\documentclass[prl,twocolumn,superscriptaddress,showpacs,amsmath,amssymb]{revtex4-1}
\usepackage[dvipdfmx]{graphicx}
\usepackage[dvipdfmx]{color}
\usepackage{subfigure}
\usepackage{natbib}
\usepackage{bm}
\usepackage{siunitx}
\def\U#1{{\rm #1}} 
\def\u#1{_{\rm #1}}
\newcommand{\ket}[1]{| #1 \rangle}
\newcommand{\bra}[1]{\langle #1 |}
\def\tr{\U{tr}}

\begin{document}
\title{
  A cavity-enhanced broadband photonic Rabi oscillation 
}
\author{Rikizo Ikuta}
\affiliation{Graduate School of Engineering Science, Osaka University,
  Toyonaka, Osaka 560-8531, Japan}
\affiliation{
Center for Quantum Information and Quantum Biology, 
Osaka University, Osaka 560-8531, Japan}
\author{Toshiki Kobayashi}
\affiliation{Graduate School of Engineering Science, Osaka University,
  Toyonaka, Osaka 560-8531, Japan}
\affiliation{
Center for Quantum Information and Quantum Biology, 
Osaka University, Osaka 560-8531, Japan}
\author{Tomohiro Yamazaki}
\affiliation{Graduate School of Engineering Science, Osaka University,
Toyonaka, Osaka 560-8531, Japan}
\author{Nobuyuki Imoto}
\affiliation{
Center for Quantum Information and Quantum Biology, 
Osaka University, Osaka 560-8531, Japan}
\author{Takashi Yamamoto}
\affiliation{Graduate School of Engineering Science, Osaka University,
Toyonaka, Osaka 560-8531, Japan}
\affiliation{
Center for Quantum Information and Quantum Biology, 
Osaka University, Osaka 560-8531, Japan}

\begin{abstract}
  A coherent coupling among different energy photons
  provided by nonlinear optical interaction 
  is regarded as a photonic version of the Rabi oscillation.
  Cavity enhancement of the nonlinearity reduces energy requirement significantly 
  and pushes the scalability of the frequency-encoded photonic circuit 
  based on the photonic Rabi oscillation.
  However, confinement of the photons in the cavity severely limits
  the number of interactable frequency modes. 
  Here we demonstrate a wide-bandwidth and efficient photonic Rabi oscillation 
  achieving full-cycle oscillation based on a cavity-enhanced nonlinear optical interaction
  with a monolithic integration.
  We also show its versatile manipulation beyond the frequency degree of freedom 
  such as an all-optical control for polarizing photons with geometric phase. 
  Our results will open up full control accessible to synthetic dimensional photonic systems 
  over wide frequency modes as well as a large-scale photonic quantum information processing.
\end{abstract}
\maketitle

\begin{figure*}
 \begin{center}
      \scalebox{1}{\includegraphics{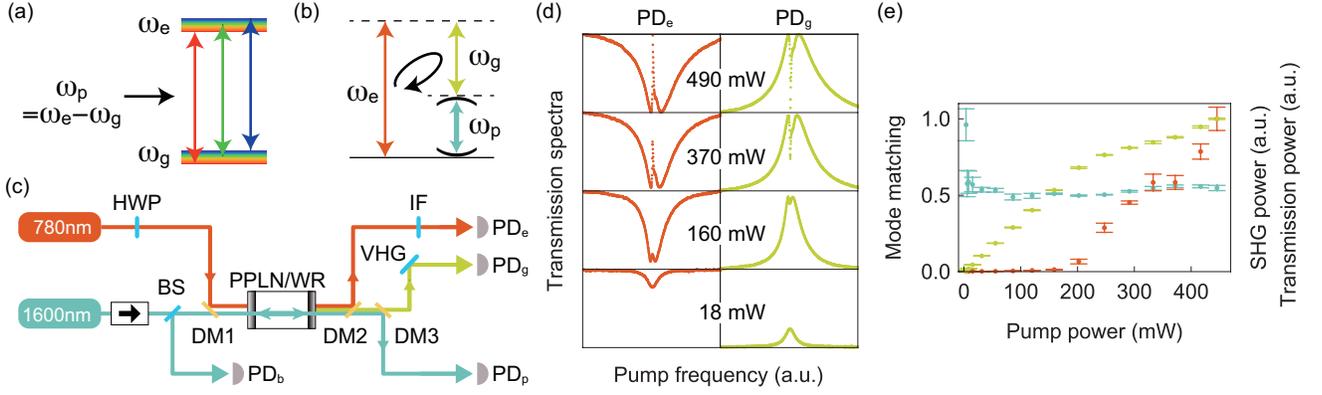}}
      \caption{
    (a)
      Two-level photonic systems coherently-coupled by photonic Rabi oscillation.
      Due to the virtual level structure, 
      the ground and excited levels are regarded as continuously distributed. 
      (b)
      Energy diagram of $\chi^{(2)}$ interaction
      equivalent to the coherent two-level systems. 
      The pump light at angular frequency $\omega\u{p}$ is confined in the cavity. 
      (c)
      Experimental setup. 
      (d)
      Pump power dependencies of
      the transmission spectra of \SI{780}{nm} light and \SI{1522}{nm} light.
      The power in the figure is the pump power measured in front of the PPLN/WR. 
      (e)
      Input pump power dependencies of 
      the mode matching~(coupling efficiency) of the pump light 
      to the PPLN/WR~(purple, left-hand axis),
      and SHG/transmission power of the pump light~(red/brown, right-hand axis). 
      The error bars indicate one standard deviation.
      For estimating the mode matching, 
      we used the data with the standard deviations smaller than \SI{5}{\%}~(The leftest four data are omitted).}
 \label{fig:setup}
 \end{center}
\end{figure*}

Rabi oscillation, which is a cyclic rotation 
between coherently-coupled two atomic levels driven by an optical field, 
is one of the most fundamental building block in atomic physics, 
and has been used for numerious technologies 
such as atomic clocks~\cite{Mcgrew2018}, sensors~\cite{Bongs2019},
quantum communication~\cite{Kimble2008} and computing~\cite{Cirac2000,Madjarov2020}.
In a photonic system, the coherent two-level system is implemented 
by two distinct photonic frequencies 
coupled by nonlinear optical interaction with a pump light. 
Recently, such a photonic Rabi oscillation has been applied to 
a single photon known as quantum frequency conversion~(QFC)~\cite{Kumar1990} 
in quantum information processing,
which can create a coherent superposition of the frequency modes in the single photon. 
Apart from atomic systems using naturally-determined energy levels, 
the photonic systems use virtual levels determined 
by the pump frequency corresponding to the difference
of the two frequencies~(Fig.~\ref{fig:setup}~(a)). 
This feature allows for interaction of densely-embedded optical frequency modes
over a wide range, 
and offers flexible manipulation of photonic frequencies
that form the inherently-equipped high dimensional Hilbert space. 

So far, several experimental demonstrations of quantum operations 
on the frequency-encoded qubits based on optical nonlinearities 
have been performed~\cite{Kobayashi2016,Clemmen2016,Joshi2020}
aiming at frequency-domain photonic quantum information processing
such as universal quantum computation based on multi-stage 
nonlinear optical interaction with the photonic Rabi oscillation~\cite{Niu2018,Niu2018-2,Krastanov2020}. 
Typically, the full Rabi cycle in $\chi^{(2)}$-based QFC 
corresponding to the so-called $2\pi$ pulse 
needs over a watt-class continuous wave~(cw) pump power 
even when waveguided crystals are used~\cite{Ikuta2011,Albrecht2014,Ikuta2018,Bock2018,Dreau2018,Yu2020,Leent2020}. 
Much more pump power is required for QFCs with 
bulk crystals or $\chi^{(3)}$ media due to
shorter interaction time or smaller nonlinearity. 
This prevents the scalable integration of the photonic Rabi osillation 
for simulating more complex quantum systems. 
A promising approach for saving the cw pump power is 
an enhancement of optical nonlinearity 
by using an optical cavity. 
However, cavity systems in which all relevant lights are confined 
such as ring resonator systems~\cite{Guo2016,Lu2019} 
severely limit accessible frequencies and bandwidths of the photons,
which sacrifices the feature of the photonic Rabi oscillation. 

In this study, without losing the acceptable frequencies and bandwidth, 
we demonstrate cavity enhancement of the photonic Rabi oscillation. 
We utilize 
a periodically-poled lithium niobate~(PPLN) waveguide as a $\chi^{(2)}$ medium 
with a cavity {\it only for} the pump light~(Fig.~\ref{fig:setup}~(b)), 
which we call the PPLN waveguide resonator~(PPLN/WR) hereafter.
The internal enhancement factor is estimated to be over 10 compared with convential QFCs 
with achieving a maximum transition probability over \SI{90}{\%}.
Such an efficient photonic Rabi oscillation realizes 
full cycle of the coherent rotation between the two frequency modes for the first time. 
This also enables an all-optical control of polarizing photons
beyond the manipulation of frequency degree of freedom~(DOF). 
Thanks to the broad bandwidth property over \SI{100}{GHz}
of the PPLN waveguide~\cite{Ikuta2011},
we can simultaneously perform the operation over several dozen frequency and polarizing modes 
forming hyper-entangled states on dense quantum frequency combs~\cite{Kues2019,Ikuta2019}. 

We explain the quantum theory of $\chi^{(2)}$-based photonic Rabi oscillation 
between angular frequencies $\omega\u{g}$ and $\omega\u{e}$ 
driven by a pump light at an angular frequency $\omega\u{p}(=\omega\u{e}-\omega\u{g})$ 
for the nonlinear optical interaction~\cite{Ikuta2011}. 
The subscripts of the two energy levels `e' and `g'  mean `excited' and `ground' frequencies. 
The energy level relevant to the process is shown in Fig.~\ref{fig:setup}~(b). 
When the pump light is sufficiently strong,
and the perfect phase matching condition is satisfied, 
the effective Hamiltonian is described by
$H=i\hbar (g^*a^\dagger\u{e}a\u{g} - ga\u{e}a^\dagger\u{g})$, 
where $a\u{e(g)}$ is an annihilation operator 
of the higher~(lower) frequency mode. 
The effective coupling contant $g=|g|e^{i\phi}$ is proportional to
the complex amplitude of the pump light with phase $\phi$. 
From this Hamiltonian,
annihilation operators $a\u{e,out}$ and $a\u{g,out}$ 
for the higher and lower frequency modes from the nonlinear optical medium
are described by
\begin{eqnarray}
\left[
  \begin{array}{c}
    a\u{e,out}\\
    a\u{g,out}
  \end{array}
  \right]
  =
  \left[
  \begin{array}{cc}
    \cos\frac{\theta}{2} & - e^{i\phi}\sin\frac{\theta}{2} \\
 e^{-i\phi}\sin\frac{\theta}{2}  &  \cos\frac{\theta}{2}
  \end{array}
\right]
\left[
\begin{array}{c}
    a\u{e}\\
    a\u{g}
  \end{array}
  \right],
  \label{eq:evolution}
\end{eqnarray}
where $\theta/2 = |g|\tau$, and 
$\tau$ is the interaction time of the light through the nonlinear optical medium.
This process is equivalent to the atomic Rabi oscillation 
driven by the external optical field resonant with the two energy levels. 
The branching ratio 
of the transition matrix can be adjusted by the pump power.
We notice that at $\theta = 2 \pi$,
the light at the initial frequency mode is obtained with a unit probability, 
while $\pi$ phase shift understood by the geometric phase is added on its complex amplitude. 
When the nonlinear optical medium is placed inside a cavity confining only the pump light,
the coupling strength $|g|$ is enhanced without any other modifications, 
resulting in a higher Rabi frequency
while preserving the intrinsic bandwidth of the nonlinear interaction. 

The experimental setup for the photonic Rabi oscillation
between \SI{780}{nm} and \SI{1522}{nm} light driven
by the cavity-enhanced pump light at \SI{1600}{nm} 
is shown in Fig.~\ref{fig:setup}~(c).
The cw pump light and the \SI{780}{nm} light with a power of \SI{1}{mW} 
are combined at a dichroic mirror~(DM1), and then focused on the PPLN/WR.

The PPLN waveguide used in our experiment has 
a periodically-poling period
for satisfying the type-0 quasi-phase-matching condition, 
and the polarization of the light relevant to the Rabi oscillation is V. 
The length of the waveguide is 20~mm.
For forming the singly-resonant PPLN waveguide resonator
with the Fabry-P\'{e}rot structure, 
the end faces of the waveguide are flat polished, 
and coated by dielectric multilayers.
The reflectance for \SI{1600}{nm} is about \SI{98}{\%}, 
and the quality factor of the cavity for the pump light 
is about $3.2\times 10^6$~\cite{Ikuta2019}. 
For \SI{780}{nm} and \SI{1522}{nm},
anti-reflective coatings are achieved
with the reflectances of \SI{5}{\%} and \SI{0.1}{\%}, respectively. 
The coupling efficiency of \SI{780}{nm} light to the PPLN/WR is 0.91. 

After the PPLN/WR, 
the \SI{780}{nm} light is reflected by DM2,
and passes through an interference filter~(IF) with a bandwidth of \SI{3}{nm} 
followed by a photo detector~($\U{PD}\u{e}$). 
The \SI{1522}{nm} light and the pump light pass through DM2,
and they are separated by DM3. 
The \SI{1522}{nm} light passing through DM3 is diffracted 
at a volume holographic grating~(VHG) with a bandwidth of \SI{1}{nm}
and is detected by $\U{PD}\u{g}$. 
The pump light reflected at DM3 is detected by $\U{PD}\u{p}$. 
The pump light coming back from the PPLN/WR is monitored by $\U{PD}\u{b}$. 

The branching ratio between the two frequency modes
characterized by rotation angle $\theta$ of Rabi oscillation in Eq.~(\ref{eq:evolution}) 
was measured by the PDs for various pump powers
while scanning the pump frequency. 
Examples of the observed power spectra for \SI{780}{nm} and \SI{1522}{nm} light 
are shown in Fig.~\ref{fig:setup}~(d).
We see that when the pump light was resonant to the PPLN/WR,
the transition process was observed as a dip and a peak of the power spectra. 
For the pump power satisfying $0\leq \theta\leq \pi$, 
a higher pump power leads to the deeper dip and the higher peak of the spectra~(the bottom figure). 
For the pump power such that $\theta > \pi$ is satisfied~(the other three figures), 
the dip at \SI{780}{nm} and the peak at \SI{1522}{nm}
were respectively turned into upward and downward. 
This behavior shows the frequency recovery 
from \SI{1522}{nm} to \SI{780}{nm} 
after the transition from \SI{780}{nm} to \SI{1522}{nm}. 

\begin{figure}
 \begin{center}
    \scalebox{1.}{\includegraphics{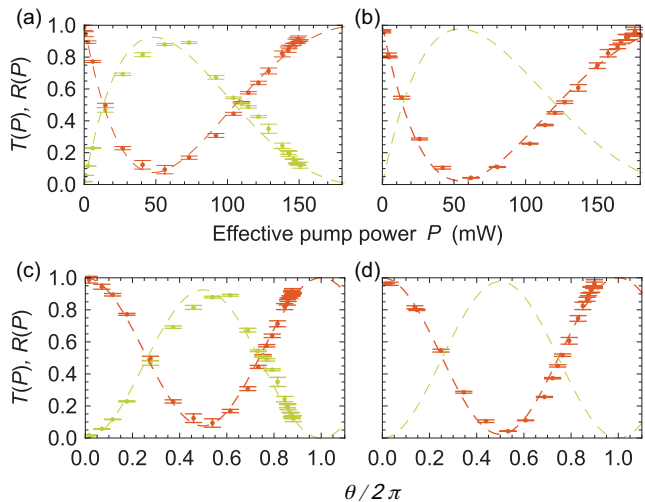}}
    \caption{
      (a) and (b) are for the signal light 
      coming from the direction same as and opposite to the pump light.
      (c) and (d) show
      dependency on $\theta$ in the Rabi cycle instead of the effective pump power
      in (a) and (b). 
    }
 \label{fig:conv}
 \end{center}
\end{figure}
The pump power was measured just before the PPLN/WR. 
In order to estimate an effective pump power $P$ used for the Rabi oscillation, 
we measured an optical mode matching~(a coupling efficiency) to the resonator
of the pump light from the reflection spectra at $\U{PD\u{b}}$, 
and second harmonic generation~(SHG) at \SI{800}{nm} of the pump light at $\U{PD\u{e}}$ 
as an unexpected nonlinear optical interaction.
We show the experimental results in Fig.~\ref{fig:setup}~(e).
From the figure, 
we estimated the amount of the mode matching to be 0.52 on average. 
The second harmonic light at \SI{800}{nm} was measured by $\U{PD}\u{e}$
without the \SI{780}{nm} signal light and IF. 
In Fig.~\ref{fig:setup}~(e), we see that SHG power is gradually increased. 
Corresponding to this, the transmitted pump power measured by $\U{PD}\u{p}$ 
begins to deviate from the proportional relationship with the input pump power. 
This indicates that the coupled pump power is consumed by SHG, 
and the transmitted power reflects the remaining pump power $P$ used for the frequency transition. 
From the experimental data for $< \SI{200}{mW}$ input pump power,
in which SHG is negligibly small, we estimated the conversion factor 
between the transmitted pump power and $P$ with considering the mode matching 0.52, 
and then we determined $P$ for all input pump power. 

Together with the estimated values of $P$ 
and the observed spectra of \SI{780}{nm} and \SI{1522}{nm} light, 
we plot the staying probability~(transmittance) $T(P)$
and the transition probability~(reflectance) $R(P)(:=1-T(P))$
in Figs.~\ref{fig:conv}~(a) and (c)
based on the normalization method in Refs.~\cite{Ikuta2013-3,Kobayashi2016}. 
The maximum transition probability
corresponding to the $\pi$ pulse was achieved at $\sim$ \SI{50}{mW} pump power. 
The best fit to $T(P)$ with a function $1 - A\sin^2(\sqrt{BP}L)$
gives $A=0.92$ and $B=\SI{0.13}{W^{-1} mm^{-2}}$, 
where $L=$\SI{20}{mm} is the waveguide length.
The observed coupling constant $B$ is over 10 times larger than
previously-reported values in QFC experiments without cavity systems~\cite{Ikuta2011,Albrecht2014,Ikuta2018,Bock2018,Dreau2018,Yu2020,Leent2020}. 
As a result, the cavity enhancement effect was clearly observed.
We will give the detailed discussion about the enhancement later. 

The pump light confinement in the Fabry-P\'{e}rot cavity 
allows for the Rabi oscillation regardless of the input direction of the pump light. 
To see this, 
we swapped the positions of the \SI{780}{nm} light source and the detector $\U{PD}\u{e}$. 
The experimental result is shown in Fig.~\ref{fig:conv}~(b) and (d). 
As is expected, the oscillation of the frequency transition was surely observed. 
The best fit to $T(P)$ with $1 - A' \sin^2(\sqrt{B' P}L)$
gives $A'=0.98$ and $B'=\SI{0.11}{W^{-1} mm^{-2}}$. 
These estimated values are similar to $A$ and $B$,
and the bidirectional photonic Rabi oscillation was successfully achieved. 

To see the amount of the cavity enhancement, 
we compare the value of the normalized coupling constant $B=\SI{0.13}{W^{-1}mm^{-2}}$ 
with the values in previous reported papers. 
In Ref.~\cite{Ikuta2011}, 
the QFC system including the PPLN waveguide and wavelengths of the relevant light 
is almost the same as this experimental setup except for the cavity coating. 
The value of the coupling constant is calculated to be $\SI{0.009}{W^{-1} mm^{-2}}$ 
from the reported experimental data. 
From these results, 
an enhancement factor of the coupling constant is estimated to be 13. 
While the value is slightly smaller than
a theoretically predicted value~\cite{Rogener1988,Stefszky2018} 
$F/\pi\sim 19$ with finesse $F=59$~\cite{Ikuta2019}, 
our result surely showed that the coupling strength becomes an order of magnitude larger. 
This statement holds in comparison to other QFC experiments
using PPLN waveguides without cavities in various situations~\cite{Albrecht2014,Ikuta2018,Bock2018,Dreau2018,Yu2020,Leent2020} 
in which the coupling constant from $0.003$ to $\SI{0.012}{W^{-1} mm^{-2}}$ were observed. 

\begin{figure}
 \begin{center}
    \scalebox{1.}{\includegraphics{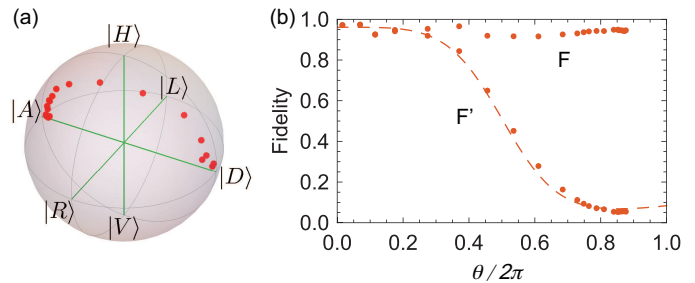}}
    \caption{
      (a)
      The Poincar\'{e} sphere~(Bloch sphere) for the polarization state of light. 
      $\ket{R}$~($\ket{L}$) is the right~(left) circular polarization state.  
      The trajectory of the states from $\theta=0$ to $\theta\sim \pi$
      is represented by red circles from near $\ket{D}$ to $\ket{A}$.
      (b)
      $\theta (P)$ dependencies of fidelities $F$ and $F'$.
      The dotted curve is theoretically obtained by using experimental parameters. 
    }
 \label{fig:bloch}
 \end{center}
\end{figure}
The photonic Rabi oscillation can be used for the polarization rotation of the signal light. 
The experiment for such the operation was performed by the same setup in Fig.~\ref{fig:setup}~(c). 
The input signal light at \SI{780}{nm} is set to diagonal polarization 
whose state is written as $\ket{D}:=(\ket{H}+\ket{V})/\sqrt{2}$, 
where $\ket{H}$ and $\ket{V}$ are states 
for the horizontal and vertical polarization of light, respectively.
Because only the V-polarized light is frequency converted by the type-0 quasi phase-matched PPLN/WR, 
the polarization state of the light at \SI{780}{nm} is transformed into 
$\ket{\psi_{\theta}}:={\mathcal N}(\ket{H}+\cos\frac{\theta}{2}\ket{V})$ ideally,
where ${\mathcal N}:=(1+\cos^2\frac{\theta}{2})^{-1/2}$ is the normalization constant. 
For \SI{780}{nm} light coming from the PPLN/WR, 
the polarization state tomography was performed 
by inserting a polarization analyzer composed of 
a quarter wave plate, a HWP and a polarizing BS
into the optical path just before $\U{PD}\u{e}$. 

In Fig.~\ref{fig:bloch}~(a), the trajectory of the polarization states
on Poincar\'{e} sphere~(or Bloch sphere) estimated from Stokes parameters is shown. 
As the pump power increases, 
the initial state near $\ket{\psi_0}(=\ket{D})$ is transformed towards 
$\ket{\psi_{2\pi}}=\ket{A}:=(\ket{H}-\ket{V})/\sqrt{2}$, 
via a state near $\ket{\psi_{\pi}}=\ket{H}$. 
For quantitative evaluation of the rotation, 
we borrowed the density operator representation $\rho_{\theta}$ 
of the polarization state for various values of $\theta=\theta(P)$ 
from the quantum information theory. 
We evaluated the fidelities $F:=\bra{\psi_\theta}\rho_{\theta}\ket{\psi_\theta}$ 
of $\rho_{\theta}$ to the ideal state $\ket{\psi_\theta}$ 
and $F':=\bra{\psi_0}\rho_{\theta}\ket{\psi_0}$ 
of the states with turning on and off the pump light. 
The results are shown in Fig.~\ref{fig:bloch}~(b). 
For every $\theta$, $F$ is higher than 0.9, 
and thus the polarization state is successfully rotated 
by the Rabi oscillation for any pump power with a resonant frequency. 
The result of $F'$, which approaches zero as $\theta$ increases, also shows the rotation effect.
We notice that for $\theta=2\pi$, 
$F'$ is close to zero 
due to the effect of the geometric phase added 
through the full Rabi cycle ~\cite{Vandevender2007,Langford2011,Karnieli2019}. 
The above polarization rotation was controlled 
by amplitude modulation~(AM) of the pump light at a frequency resonant on the cavity. 
In addition to the AM control, 
the cavity structure enables us to toggle the polarization rotation 
by frequency modulation~(FM) of the pump light 
between off-resonant and on-resonant condition with a fixed pump power. 

We construct a theoretical model for the state transformation 
by the photonic Rabi oscillation. 
The observed maximum transition efficiency is $A=0.95$. 
We assume that 
the imperfection is originated from the propagation mode mismatch 
of the signal and the pump light. 
We denote the propagation mode which interacts with the pump light by $\ket{x}$
and its orthogonal mode by $\ket{\bar{x}}$. 
Under the assumption, 
when the input state is a pure state with diagonal polarization, 
the initial state to QFC can be written as 
$\ket{\psi\u{in}}:=\ket{D}(\sqrt{A}\ket{x}+\sqrt{1-A}\ket{\bar{x}})$. 
After the Rabi oscillation with $\theta$, 
the normalized output state $\ket{\psi\u{out}}$ is ideally described by 
\begin{eqnarray}
  \hspace{-10pt}
  \ket{\psi\u{out}}:={\mathcal N_0}\left(
  \sqrt{A}{\mathcal N}^{-1}\ket{\psi_{\theta}}\ket{x}
  +\sqrt{2(1-A)} \ket{D}\ket{\bar{x}}\right), 
\end{eqnarray}
where 
${\mathcal N_0}:=(A{\mathcal N}^{-2}+2(1-A))^{-1/2}$ is a normalization factor. 
In our experiment, the reconstructed states $\rho_\theta$ were slightly impure,
namely  $\tr(\rho_\theta^2) < 1$. 
We model the output state including the impurity as
$\rho\u{th}:=p \ket{\psi\u{out}}\bra{\psi\u{out}}+(1-p) I/2$,
where $p:=(2\langle \tr(\rho_\theta^2)\rangle -1)^{1/2}$ and $I$ is the identity operator. 
By using experimentally observed values
$A=0.95$ and $\langle \tr(\rho_\theta^2)\rangle = 0.93$, 
we obtained the fidelity $F\u{th}:=\bra{\psi\u{in}}\rho\u{th}\ket{\psi\u{in}}$
as shown in Fig.~\ref{fig:bloch}~(b). 
We see that the experimental results are in good agreement with 
the curve theoretically predicted with the use of the experimental parameters. 

In our demonstration for the polarization state rotation, 
we measured only the polarization state of \SI{780}{nm} light. 
When we consider the frequency DOF
composed of \SI{780}{nm} and \SI{1522}{nm} modes, 
which we denote by $\ket{\omega\u{h}}$ and $\ket{\omega\u{l}}$, 
the photonic Rabi oscillation system works on 
a single-photon two-qubit state~\cite{Kim2003} 
formed by the polarization and the frequency modes of the single photon.
Because the PPLN/WR has polarization dependency, 
an input state $\ket{D}\ket{\omega\u{h}}$ is deterministically transformed 
into the single-photon Bell states as 
$(\ket{H}\ket{\omega\u{h}}\pm \ket{V}\ket{\omega\u{l}})/\sqrt{2}$ 
by the Rabi oscillation with $\theta=\pi/4$ and $3\pi/4$.
These operations correspond to the controlled NOT gates.
Combining frequency dependent WPs, the other two Bell states can be easily generated. 
The measurement of the four Bell states is also deterministically achieved 
by the same setup. 

In conclusion,
we have demonstrated cavity enhancement of photonic Rabi oscillation 
while keeping the flexible choice of frequencies and wide acceptance bandwidths of photons. 
The enhancement of the nonlinear optical coupling was 
10 times larger than convential frequency converters. 
This leads to the observation of the full Rabi cycle between the two photonic frequencies, 
resulting in all-optical versatile manipulation beyond on the frequency DOF. 
Our results will open up
a large-scale photonic quantum information processing based on frequency modes
including hyper-entangled systems~\cite{Barreiro2008,Reimer2019}. 
Furthermore, considering more than two virtual energy levels and multiple pump lasers, 
the photonic system will enable to simulate and design the complicated atomic systems. 

RI, NI and TY acknowledge members of Quantum Internet Task Force (QITF) for comprehensive and interdisciplinary discussions of the Quantum Internet.
This work was supported 
by CREST, JST JPMJCR1671; MEXT/JSPS KAKENHI Grant Number 
JP20H01839 and JP18K13483; 
Asahi Glass Foundation. 

\end{document}